\documentclass[aps,prl,color,twocolumn,showpacs]{revtex4}
\usepackage{graphicx}
\usepackage{color}
\begin{document}

\title{Depletion effects and loop formation in self-avoiding polymers}
\author{N.M. Toan$^{1}$, D. Marenduzzo$^2$,
P.R. Cook$^3$ and C. Micheletti$^{1}$,}
\affiliation{$^1$ International School for Advanced Studies (SISSA)
  and CNR-INFM, Via Beirut 2-4, 34014 Trieste, Italy \\
$^2$ SUPA, School of Physics, University of Edinburgh, Mayfield Road, Edinburgh, EH9 3JZ, Scotland\\
$^3$ Sir William Dunn School of Pathology, University of Oxford, South
  Parks Road, Oxford, OX1 3RE, England}
\begin{abstract} 
Langevin dynamics is employed to study the looping kinetics of
self-avoiding polymers both in ideal and crowded solutions. A 
rich kinetics results from the competition of two crowding-induced
effects: the depletion attraction and the enhanced viscous
friction. For short chains, the enhanced friction slows down 
looping, while, for longer chains, the
depletion attraction renders it more frequent and persistent. We
discuss the possible relevance of the findings for 
chromatin looping in living cells.
\end{abstract}
\maketitle

The kinetics and thermodynamics of the folding of a flexible polymeric
chain into a loop are central issues in polymer physics
\cite{DoiEdwards,loop,pseudoknot,Stella}. Renewed interest in this
classic problem has been fuelled by the introduction of novel
manipulation techniques \cite{dekker3C} that provide unprecedented
insight into the mechanics and flexibility of various biopolymers. In
particular, it has been shown that, in the cell nucleus, DNA regions
separated by several $\mu$m's on the genetic map can nevertheless be
in molecular contact \cite{dekker3C}.  Is such looping generated by
{\it active} mechanisms (e.g. molecular motors), or merely by {\em
passive} thermodynamic mechanisms (e.g., diffusion)? Some light can be
shed on these issues by comparing experimental observations with the
statistics and dynamics of looping predicted by general polymer
models. The systematic application of this strategy has so far been
hindered by the dependence of cyclization dynamics on many time scales
even for the simplest phantom polymer models \cite{canadian}.

Here, we go beyond the treatment of phantom chains and focus on the
impact of steric effects. We not only consider the polymer self
avoidance \cite{DoiEdwards,vologodski} but also incorporate excluded
volume interactions of the chain with a surrounding crowded
environment, treated as a collection of small monodisperse globular
particles (microspheres) which induce an entropic attraction on larger
objects in solution\cite{AO}. To date, the investigation of this
intriguing {\it depletion effect}, has been mainly studied in
polydispersed colloidal solutions \cite{AO,yodh}. Understanding how
crowding affects the behaviour of a single self-avoiding polymer thus
represents a novel and important topic in macromolecular physics.  It
also has immediate implications in cell and systems biology, as cells
are so crowded with globular proteins and RNAs
\cite{Thirumalai2005,Kamien2005,Peter,Minton}.  We show that crowding
affects the occurrence and persistence of loops in self-avoiding
polymers in diverse ways, according to the length of the polymer chain
and the size of the constitutive monomers.  Besides uncovering new
physics, our results may be relevant to the understanding of chromatin
looping {\it in vivo}.  Specifically, we will discuss whether and to
what extent the depletion attraction may explain existing observations
in cell biology that active polymerases, attached to chromatin,
cluster into supramolecular ``factories'' of up to $\mu$m size during
transcription and replication \cite{Peter}.

\begin{figure}[t]
\includegraphics[width=1.6in]{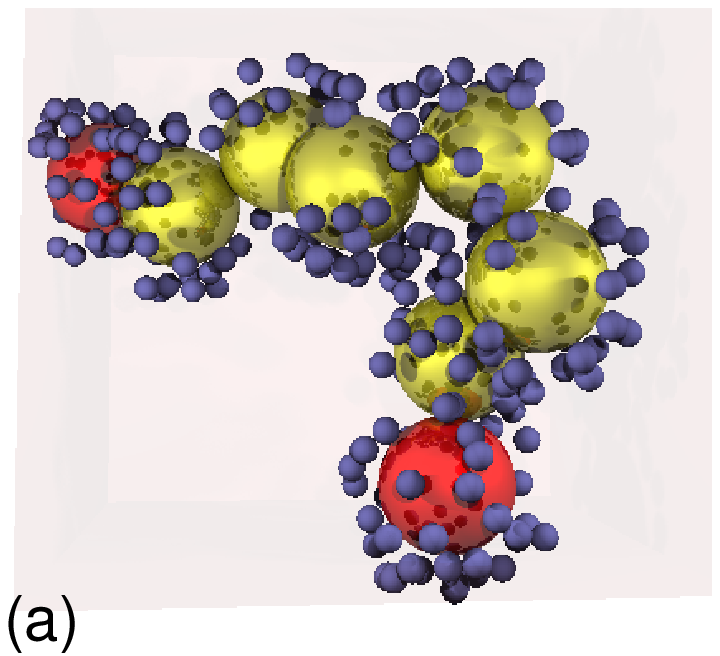}
\includegraphics[width=1.6in]{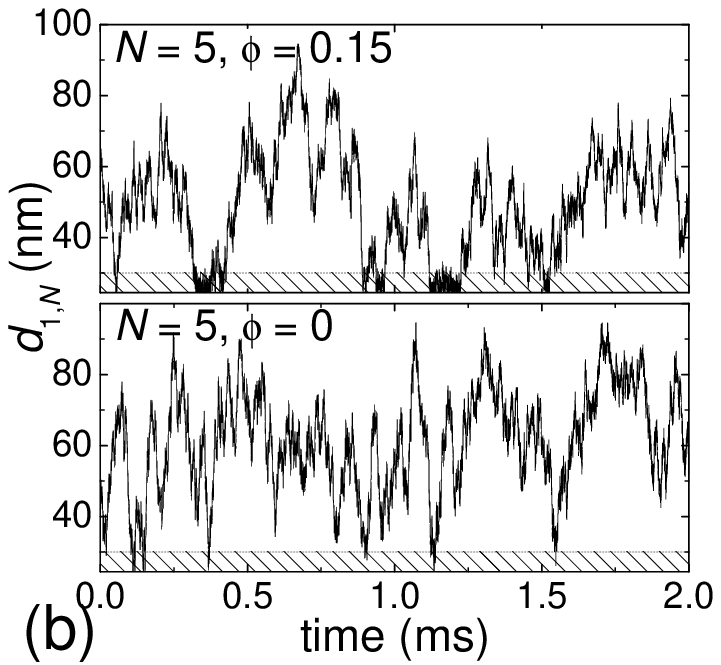}\\
\includegraphics[width=1.6in]{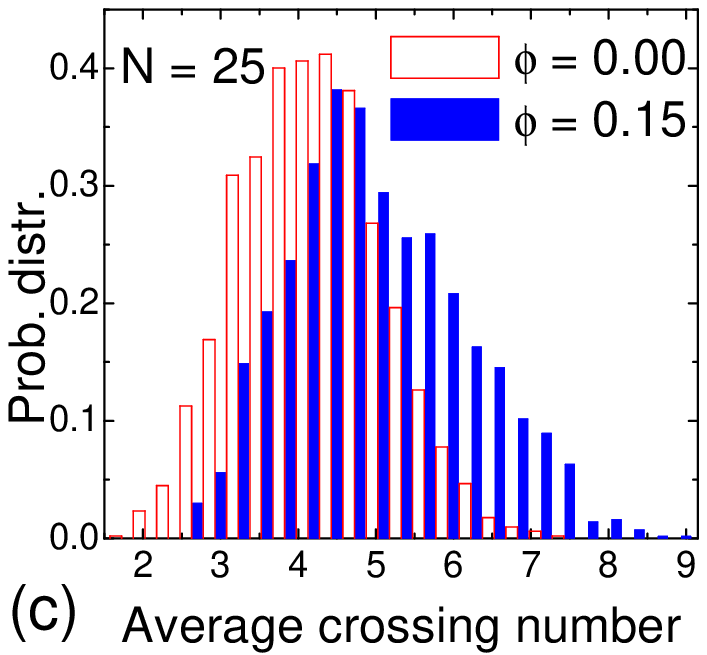}
\includegraphics[width=1.6in]{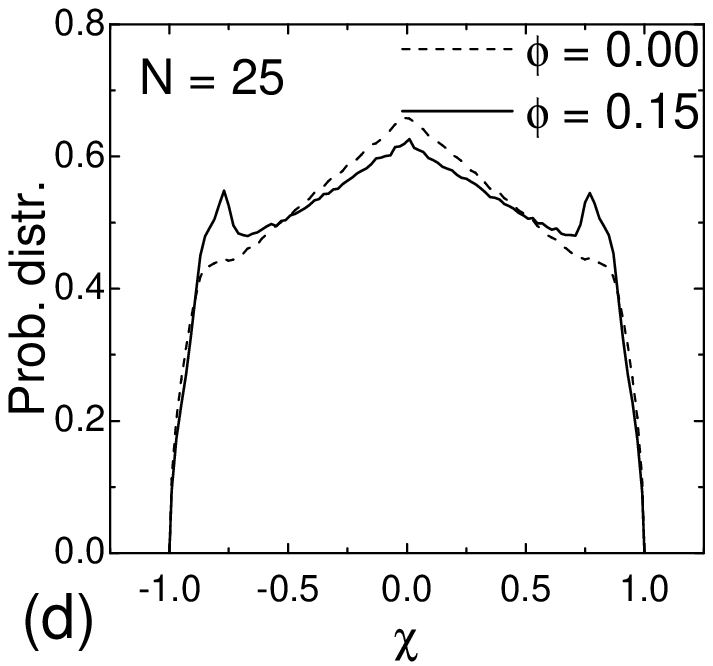}
\caption{[Color online] (a) Sketch of a chain of $N=8$ beads (end
spheres highlighted) of radius $R=12.5 $nm. Microspheres ($r=2.5 $nm
and $\phi=0.15$) within 7.5 nm of the chain surface are shown. (b)
Time evolution of the end-to-end distance, $d_{1,N}$, in chains of
$N=5$ beads subject to the effective depletion potential accounting
for the presence ($\phi=0.15$) and absence ($\phi=0$) of the
microspheres. Looping occurs when $d_{1,N} < 2(R+r) =30$ nm (shaded
area). Probability distribution of (c) average crossing number and (d)
chirality for $\phi=0$ and $\phi=0.15$.}
\label{fig:model}
\end{figure}

A traditional string-and-beads model will be used to describe polymer
chains.  We consider two simple cases: a plain self-avoiding string of
$N$ equally-sized spherical beads of radii $R_{i=1,...,N}$=12.5nm, and
one with larger beads at the ends. This sphere size was chosen to
relate our polymer to an eukaryotic chromatin fiber, whose effective
diameter and persistence length are both $\sim$ 25 nm
\cite{chromatin}.  The case of larger end beads is instead motivated
by the study of chromatin loops with an attached genome-grabbing
machinery (a large transcription or a replication complex which
locally increases the effective fiber diameter \cite{Peter}).  The
potential energy, when the centres of beads $i$ and $j$ are at a
distance $d_{i,j}$, is
\begin{equation}
V_{c} = \epsilon_1 \sum_{i<j} e^{-a (d_{i,j} - d^0_{ij})} -
\epsilon_2 \sum_{i} \ln [ 1 - ({d_{i,i+1} \over 1.5\ d^0_{i,i+1}})^2]
\label{eqn:Vc}
\end{equation}
\noindent where $\epsilon_1$ and $\epsilon_2$ are respectively 0.24 and
70 units of thermal energy, $\kappa_B T$, $a = 4$ nm$^{-1}$, and
$d^0_{i,j} = R_i + R_j$ is the contact distance of beads $i$. The
first term in eqn.  (\ref{eqn:Vc}) enforces the hard-core repulsion
for contacting pairs, while the second provides an attraction between
consecutive beads on the chain. Interplay between the two terms
produces a self-avoiding FENE chain \cite{kremer1} where, at
temperature $T=300$ K, the distance between consecutive beads
fluctuates by about 0.5 nm around 25 nm. The microspheres have radius
$r=2.5$ nm and occupy a fraction $\phi=0.15$ of the total available
volume. These values conservatively reflect the crowding of the
cellular environment mostly due to RNA and proteins\cite{Peter}.
Because the value of $\phi$ considered here is moderate (see 
Fig.\ref{fig:model}a), we can resort to the approximate Asakura-Oosawa
(AO) treatment \cite{AO}, which {\em does not} require to simulating
explicitly the dynamics of microspheres. More precisely, in addition
to the term of eqn.  (\ref{eqn:Vc}), the polymer is subject to the
following effective interaction potential:
\begin{eqnarray}
V_{AO} = -\frac{\phi k_BT}{16r^3}\sum_{i<j} \left(2\tilde{d}_{ij} +
3d_{ij} - \frac{3\Delta_{ij}^2}{d_{ij}} \right)\ \tilde{d}_{ij}^2\ \Theta[\tilde{d}_{ij}]
\label{eqn:ao}
\end{eqnarray}
where $\tilde{d}_{ij} = 2r+d_{i,j}^0 - d_{ij}$, $\Delta_{ij}
=\left|R_i-R_j\right|$, and the step function $\Theta$ ensures that
the AO depletion interaction vanishes at distances $>d^0 + 2r$
\cite{AO}. As $r$ is sufficiently smaller than the radii of the chain
beads it is legitimate to disregard in~(\ref{eqn:ao}) three- and many-body
interactions.  The evolution of the system, carried out for various
$N$, sizes of end spheres and values of $\phi$ was described by
overdamped Langevin dynamics
\begin{equation}
\gamma_i \dot{x}^\alpha_i = - {\partial (V_c + V_{AO} )/
\partial x^\alpha_i}\, + \xi^\alpha_i(t)\ ,
\label{eqn:lang}
\end{equation}
\noindent where $\alpha$ runs over the Cartesian components,
$\vec{x}_i$ is the position of the $i$th bead and the stochastic white
noise term obeys the fluctuation-dissipation condition: $\langle \xi
\rangle =0$, $\langle \xi_i^\alpha(t) \xi^\beta_j(t^\prime) \rangle =
2\, \delta_{\alpha,\beta} \delta_{i,j} \delta(t,t^\prime)\, k_B T\,
\gamma_i$.  The friction term $\gamma_i$ was obtained from the
Stokes-Einstein \cite{DoiEdwards} relationship: $\gamma_i =
6\pi\eta(1+2.5\phi)R_i$ where $\eta = 5$ cP\cite{Peter}. The Langevin
equation was integrated numerically by means of a predictor-corrector
scheme~\cite{AllenTildesley} and a time step of 15 ps. The viability
of eqn. (\ref{eqn:lang}) was ascertained by a preliminary successful
comparison of various dynamic and equilibrium properties with those
produced by underdamped dynamics (with masses deduced from typical
densities of biopolymers, $\rho= 1.35$ g/cm$^3$~\cite{Matthews}).

The dynamical evolution was followed starting from randomized non
self-intersecting configurations of chains with $ 3 \le N \le 30$
beads.  When investigating how looping dynamics is affected by the
size of the contacting spheres, we set $N=10$ and varied the radius of
the end spheres, $R_1=R_N$, within 12.5 and 43.75 nm. The formation of
loops was detected by monitoring the end-to-end distance, $d_{1,N}$
and comparing it to the range of the depletion attraction, $d_{1,N} <
(d^0_{1,N} + 2r)$.  To have a well-defined comparison term, the same
criterion for loop formation was adopted in the absence of crowding
effects/agents (i.e.  $\phi=0$). Typical evolutions of the end-to-end
distance are illustrated in Fig. \ref{fig:model}b.  The trajectories
were analysed to highlight how depletion interactions affect the
looping kinetics and the chains equilibrium structural properties. For
the latter issue several geometrical descriptors were considered: the
radius of gyration, virial coefficients and the distribution of local
chiralities $\chi_i = \vec{u}_{i+2,i+3}\cdot(\vec{u}_{i,i+1}\wedge
\vec{u}_{i+1,i+2})$, $\vec{u}_{i,i+1}$ being the normalised bond
vector joining residues $i$ and $i+1$. For looped configurations we
also calculated the writhe and crossing number, averaged over hundreds
of randomly oriented two-dimensional projections
\cite{dewitt}. Concerning the looping dynamics we instead characterize
the evolution of the system in terms of the mean first looping time
(MFLT) and the mean-first unlooping time (MFUT), which we establish
through the following novel procedure apt for numerical
implementation. For very long, and hence thermalised, trajectories the
MFLT is obtainable by picking at random unlooped conformations and
measuring the time to the first looping event. Configurations in a
specific ``unlooped time interval'' of duration $\tau_u$ will be
picked with weight equal to $\tau_u$ and their average first looping
time will be $\tau_u/2$. The MFLT can thus be expressed in terms of
the average duration of unlooping intervals and its second moment:
MFLT$={1\over 2}{\left<\tau_u^2\right>\over \left<\tau_u\right>}$. An
analogous formula holds for the MFUT .  The average values of MFUT and
MFLT (we have verified that the first two moments of looping and
unlooping intervals are finite) and their uncertainties were
calculated over 10 independent trajectories having maximum duration
ranging from 0.1 s for $N=3$ to $10$ s for $N=30$.
 
We first discuss the structural differences of the generated
configurations.  The short-range depletion attraction produces a
reduction of the effective size of the polymer. For the largest $N$
considered, where most conformations are unlooped, the radius of
gyration decreases by 10\% when $\phi$ goes from 0 to 0.15.  The
depletion attraction also impacts on the chain structural
organization. Indeed, the geometrical complexity of looped chains is
enhanced by crowding and the average crossing number increases with
$\phi$ (Fig. \ref{fig:model}c). This is akin to what occurs in random
rings upon compactification \cite{dewitt}.  At variance with this
case, the development of a striking trimodal character of the
chirality distribution indicates the emergence of a peculiar
structural organization (see Fig. \ref{fig:model}d). Though the chiral
biases are local, and hence do not lead to long-range structural
order, they provide qualitative support to the recent suggestion of
Snir and Kamien that depletion effects may be sufficient to drive the
formation of optimal helices in thick biopolymers \cite{Kamien2005,opthelices}.

\begin{figure}
\includegraphics[width=1.6in]{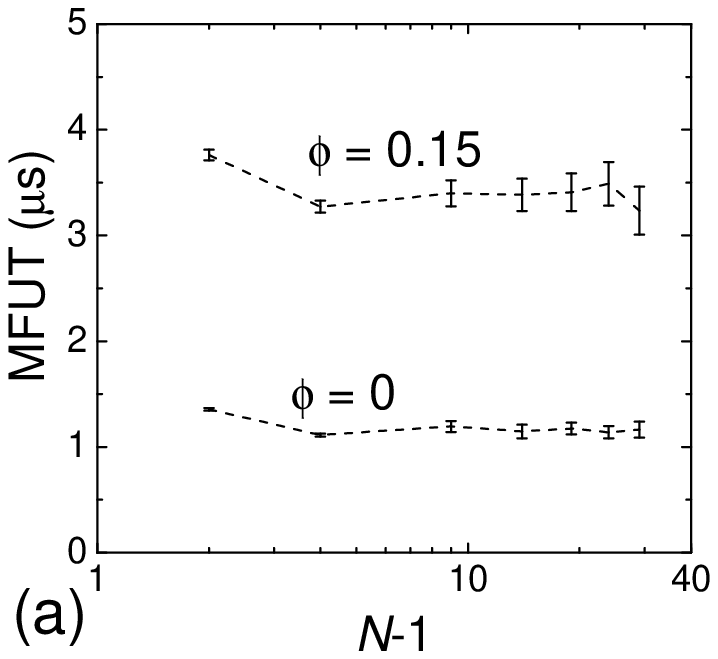}
\includegraphics[width=1.6in]{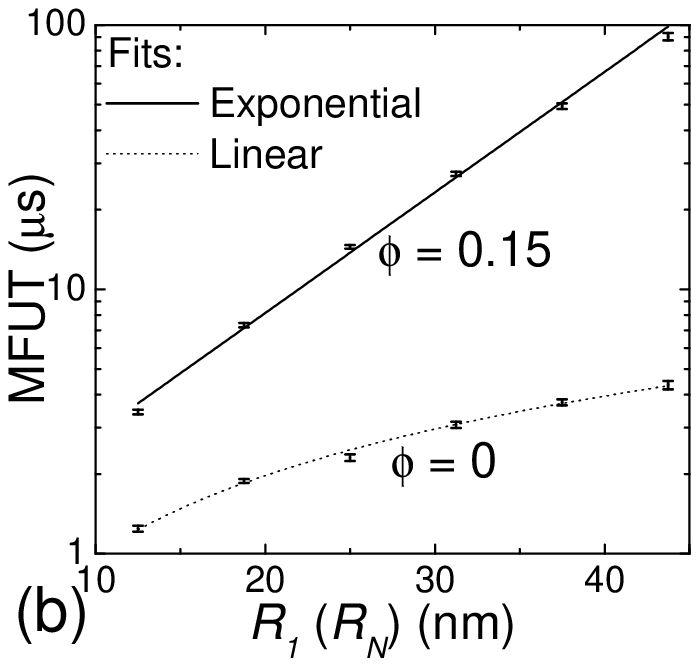}\\
\includegraphics[width=1.6in]{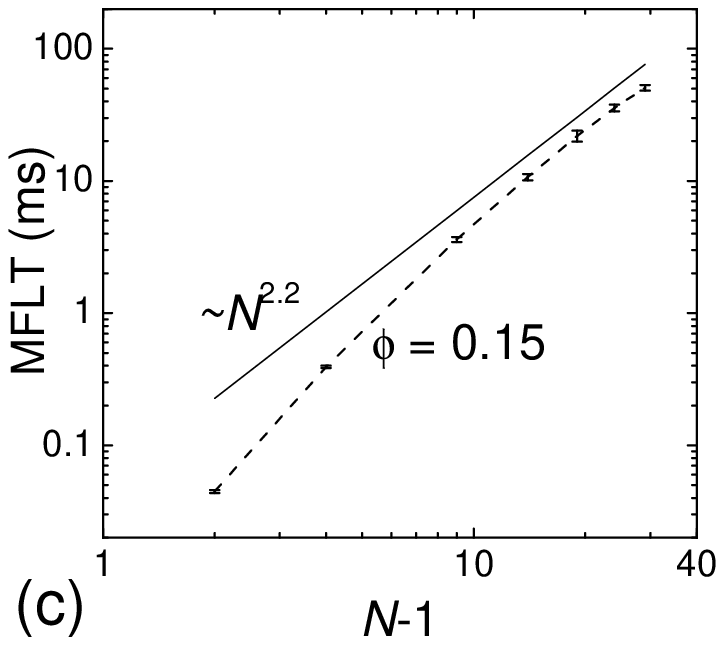}
\includegraphics[width=1.6in]{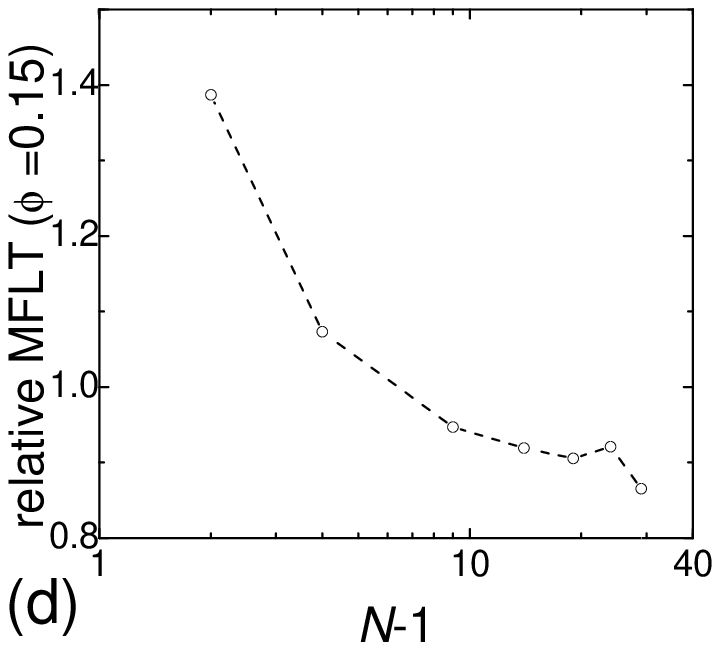}
\caption{Mean first unlooping time as a function of (a) chain length and
(b) radius of terminal spheres. The curves in (b) correspond to
exponential and linear fits of the data. (c) Mean first looping time as a function of chain length
(points and dashed line). The asymptotic law for relaxation time in
Rouse chains, $\tau_R \propto N^{2 \nu +1}$, is shown for comparison.
(d) Ratio between MFLT in the presence ($\phi=0.15$) and absence
($\phi=0$) of crowding agents. The typical error is 5\%.}
\label{fig:d}
\end{figure}

We next turn to the analysis of the looping kinetics and discuss the
behaviour of the MFUT. In the presence of microspheres, the two
contacting ends are subject to the depletion attraction and it may be
anticipated that the MFUT is larger than for $\phi = 0$. 
This expectation, qualitatively perceivable from 
Fig.  \ref{fig:model}b, is confirmed and quantified in
Fig. \ref{fig:d}a which portrays a parallel trend of unlooping time as
a function of $N$ for $\phi=0$ and $\phi=0.15$. Moderate
crowding is enough to increase the time the ends spend together
(forming a loop) by a factor of 3. 

\begin{figure}
\includegraphics[width=2.in]{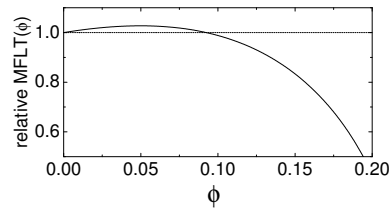}
\caption{Theoretical behaviour of the ratio between MFLT at finite
volume crowding, $\phi$, and the MFLT ($\phi=0$).}
\label{fig:toan}
\end{figure}

The behaviour of the MFLT (Fig. \ref{fig:d}c and d) is more intriguing
and harder to anticipate, owing to two opposing tendencies. On one
hand, as we saw earlier the depletion attraction reduces the effective
polymer size and hence favours looping. On the other, crowding
augments the effective viscosity of the medium, thereby slowing the
diffusive encounter of terminal beads. More precisely, the
Stokes-Einstein formula gives a 38 \% increase of friction coefficient
when $\phi=0.15$ compared to $\phi=0$.  The resulting balance between
the two opposing effects can be established by considering the
asymptotic expression for relaxation times in Rouse chains with
excluded volume. The slowest relaxation time in such chains
(assimilated to the MFLT \cite{canadian,vologodski}) increase as
$\gamma\, b^2 N^{2 \nu +1}$, where $\nu \approx 0.6$ is the scaling
exponent for self-avoiding polymers and $b$ is the effective size of
the chain monomers estimated by calculating the second virial
coefficient accounting for the depletion attraction of of eqn.
\ref{eqn:ao} \cite {DoiEdwards}. Indeed, the data collected within the
explored range of $N$, appear well compatible with this asymptotic
relationship (see Fig. \ref{fig:d}c). The Rouse scaling formula can
hence be used to quantify how the crowding-induced changes in $\gamma$
and $b$ ultimately affect the MFLT for large $N$. For the specific
case considered here, $\phi=0.15$, one finds that the MFLT is {\em
decreased} by about 17\% compared to the $\phi=0$ case. This crude
asymptotic estimate is in fair agreement with the simulation results
for $N=$15-30 which indicate that, when $\phi=0.15$, crowding
decreases the MFLT by 10 \% or more. It is interesting to consider,
within the previous approximate analysis, how the MFLT depends on
$\phi$. This information, obtained from the known functional
dependence on $\phi$ of the friction and second virial coefficients,
is shown in Fig. \ref{fig:toan} which portrays an intriguing
non-monotonic dependence: for small crowding $\phi <~ 0.1$, the
diffusive slowing dominates, while larger $\phi$ facilitate looping
via depletion-induced crumpling of the chain.

We finally discuss the dependence of looping on the size of the
terminal beads.  We considered chains of $N=10$ beads where $R_1=R_N$
were varied between 12.5 and 43.75 nm. We found that the increase in
end spheres size had minor effect on the MFLT which, for $\phi=0.15$,
changed by less than 10 \% over the explored range of terminal
radii. The near constancy of the MFLT is noteworthy since the change
of $R_1=R_N$ from 12.5 to 43.75 nm implies a two-fold increase of
probability that terminal spheres contact internal ones. In contrast,
the MFUT is greatly affected by changes in terminal radii. As
illustrated in Fig. \ref{fig:d}b for $\phi=0.15$ it increases
approximately exponentially as a function of $R_1$ and $R_N$. This
behaviour would be expected if the terminal bead exited the depletion
well (whose depth increases linearly with $R_1=R_N$) according to
simple Arrhenius kinetics. However, the increase in MFUT also reflects
the decreased diffusivity of terminal beads as a result of the linear
increase in $R$ of the friction coefficient, $\gamma$. This second
effect, dominated by the former in the presence of crowding, is
readily visible in the curve pertaining to $\phi=0$ in
Fig. \ref{fig:d}b. The dotted line represents a fit to the 6 data
points with a linear relationship in $R_1=R_N$ (with relative $\chi^2$
equal to 1.5), which would be appropriate if the diffusion coefficient
of the terminal spheres were the only factor slowing unlooping.  By
carrying out simulations for selected values of $\phi$ it was found
that increasing $\phi$ enhances the magnitude of these effects. For
example, with $\phi=0.3$, which may also be appropriate for crowding
in nucleoplasm \cite{Peter}, the MFLT is $\sim 3$ times smaller, and
the MFUT (with $R_1=R_N=25$ nm) $\sim 9$ times longer than with
$\phi=0.15$.

We now discuss the possible biological implications of our results. As
mentioned previously, a large number of active DNA (and RNA)
polymerases can attach to specific segments of a chromatin fiber thus
increasing its local thickness. Experimental observations have shown
that these scattered groups of polymerases eventually cluster into
replication (or transcription) factories thus looping the intervening
genome.  In the case of transcription factories in eukaryotes, these
structures contain $\sim 10$ or more polymerases and having size
ranging from $\sim 100$ nm up to $\sim\mu$m \cite{Peter}. Is there a
simple physical mechanism leading to their establishment?  Our study
suggests that the formation and persistence of these loops can be
aided by cellular crowding. First, using a conservative estimate of
$\phi=0.15$, we find that the depletion self-interaction of the fiber
thermodynamically facilitates looping (see e.g. Fig. \ref{fig:model}). For example,
analysis of end-to-end distances in equilibrium shows that bridging
the two ends of a 750-nm fiber, which would contain ~75 kilobases of
DNA \cite{chromatin}, costs more than 8 $k_BT$ in the absence of crowding
agents, but less than 7 $k_BT$ in their presence.  We find crowding
diminishes the looping cost by 1-2 $k_BT$ for all lengths
simulated.  All, or most, of this looping cost may be overcome by the
extra depletion attraction between the thicker ends of the loop, to
which the transcription machinery is attached, consistently with
recent theoretical predictions \cite{Peter}. Second, the results of
Fig. \ref{fig:d}d demonstrate that crowding can also aid looping {\em
dynamically} by facilitating the diffusive encounter of the ends.
This intriguing result is supported by theoretical arguments
summarised in Fig. \ref{fig:toan}, which suggest the effect is robust.
Furthermore, crowding stabilizes loops {\em once the two ends have
met}. The fact that the MFUT has an approximately exponential
dependence on the height of the depletion well (Fig. \ref{fig:d}b)
underscores the role that crowding has for the formation of
replication/transcription factories. Via eqn. (\ref{eqn:ao}) we
estimate the unlooping time of e.g. {\em two} transcription complexes
of radius $R \approx 40$ nm with $\phi \approx 0.3$ can easily exceed
0.1 s -- a macroscopic time-scale, $\sim$ the experimentally measured
persistence time of factories. In living cells the life-time of such
aggregates is probably longer because, besides other physical-chemical
factors, many complexes may come together (rather than the two
considered here) and interactions of one large bead with two or more
others cooperatively increases the stabilization due to the depletion
attraction \cite{Peter}.

In conclusion, we have considered various kinetic and thermodynamic
aspects of polymer looping in a crowded medium. The process of loop
formation is controlled by two opposing effects. On one hand looping
is entropically aided in crowded media by the depletion effect. On the
other, the enhanced friction of the medium hinders the diffusive
encounter of the chain ends. The balance of the two effects is found
to depend both on the length of the polymer chain and on the size of
the contacting ends. Specific model parameters have been used to show
quantitatively that crowding-enhanced looping formation/persistence
may be actually exploited in living cells to promote the contact of
actively replicated/transcribed chromatin regions as observed in
recent experiments \cite{Peter}. The approach outlined here
demonstrates the viability of computational and analytical approaches
to investigate the novel and stimulating problem of crowding effects
on a single polymer. It would be interesting to confront theory and
experiments on looping and unlooping times obtained from single
molecules experiments with and without crowding agents.


\end{document}